\documentclass[conference]{IEEEtran}
\IEEEoverridecommandlockouts
\usepackage{amsmath,amssymb,amsfonts}
\usepackage{algorithmic}
\usepackage{graphicx}
\usepackage{textcomp}
\usepackage{xcolor}
\usepackage{multirow}
\usepackage{booktabs}
\usepackage{todonotes}
\usepackage[noadjust]{cite}

\usepackage{soul}

\usepackage{booktabs}

\def\BibTeX{{\rm B\kern-.05em{\sc i\kern-.025em b}\kern-.08em
    T\kern-.1667em\lower.7ex\hbox{E}\kern-.125emX}}

\begin{document}

\title{Extracting Rationale for Open Source Software Development Decisions --- A Study of Python Email Archives\\
}


\author{\IEEEauthorblockN{Pankajeshwara Nand Sharma\IEEEauthorrefmark{1},
		Bastin Tony Roy Savarimuthu\IEEEauthorrefmark{2}, Nigel Stanger\IEEEauthorrefmark{3} }
	\IEEEauthorblockA{Department of Information Science,
		University of Otago\\
		Dunedin, New Zealand \\
		\IEEEauthorrefmark{1}pankaj.sharma@postgrad.otago.ac.nz,
		\IEEEauthorrefmark{2}tony.savarimuthu@otago.ac.nz,
		\IEEEauthorrefmark{3}nigel.stanger@otago.ac.nz		
		}
	}
\maketitle

\begin{abstract}
A sound Decision-Making (DM) process is key to the successful governance of software projects. In many Open Source Software Development (OSSD) communities, DM processes lie buried amongst vast amounts of publicly available data. Hidden within this data lie the rationale for decisions that led to the evolution and maintenance of software products. While there have been some efforts to extract DM processes from publicly available data, the rationale behind `how' the decisions are made have seldom been explored. Extracting the rationale for these decisions can facilitate transparency (by making them known), and also promote accountability on the part of decision-makers. This work bridges this gap by means of a large-scale study that unearths the rationale behind decisions from Python development email archives comprising about 1.5 million emails. This paper makes two main contributions. First, it makes a knowledge contribution by unearthing and presenting the rationale behind decisions made. Second, it makes a methodological contribution by presenting a heuristics-based rationale extraction system called {\em Rationale Miner} that employs multiple heuristics, and follows a data-driven, bottom-up approach to infer the rationale behind specific decisions (e.g., whether a new module is implemented based on core developer consensus or benevolent dictator’s pronouncement). Our approach can be applied to extract rationale in other OSSD communities that have similar governance structures.
\end{abstract}

\section{Introduction}\label{sec:introduction}

One of the key factors of a successful Open Source Software (OSS) project is its underlying governance mechanism. Decision-making processes are important governance artefacts that articulate how decisions are or should be made within an organization. These decision-making (DM) processes often lie hidden (i.e., implicit) within large amounts of data and hence may not be known to everyone in a transparent fashion \cite{savarimuthu2016process}. While there have been efforts to unearth these decision-making processes (\cite{keertipati2016exploring,sharma2017investigating, sharma2020mining}), the rationale depicting how certain decisions were made are not made explicit in the DM process models. 

The Merriam-Webster defines rationale as the ``the explanation of controlling principles of opinion, belief, practice, or phenomena, or an underlying reason.'' However, the literature on rationale identification and extraction in OSS design (\cite{razavian2019empirical, bhat2020evolution}), has highlighted that the focus has mainly been on the decisions themselves \cite{li2020automatic} (i.e., `what' these decisions are), not on `how' they are made. Thus, the focus of this work is extracting the rationale behind how actually these decisions are made (i.e. decision-making constructs such as \textit{consensus} and \textit{BDFL decree}). We chose Python as a case study as this OSSD community is known to follow good governance practices \cite{wang2015comparative}.

The Python language is modified and evolved by means of formal \emph{Python Enhancement Proposals (PEPs)}. There might be different rationale for accepting or rejecting a PEP. For example, the rationale behind PEP~289's acceptance was: ``Based on the favorable feedback, Guido has accepted the PEP for Py2.4''~\cite{codingforums2003b}; whereas for PEP~285, the project leader (referred to as the \emph{Benevolent Dictator For Life} or \emph{BDFL}) responded: ``Despite the negative feedback, I've decided to accept the PEP'' \cite{python2002b}. These rationale-containing sentences are hidden in email discussions inside email repositories, particularly the \textit{Python-Dev} repository of core developer discussions. 

This work aims to bridge the gap by extracting rationale for decisions made during the evolution of Python. We propose an approach called {\em Rationale Miner}, which is a heuristics-based rationale extraction system, and use this approach to extract rationale. Thus, this paper makes two key contributions: (i) a methodological contribution in the form of an approach or framework that can extract rationale that lie hidden in Python repositories, and (ii) a knowledge contribution in the form of the rationale behind decisions that were made.


This paper is organized as follows. Section~\ref{sec:background} provides a background of relevant work and also presents the research questions. Section~\ref{sec:methodology} presents the methodology used to extract rationale. Section~\ref{sec:results} presents the results, followed by a discussion of contributions in Section~\ref{sec:discussion}. Section~\ref{sec:conclusion} presents our conclusions.

\section{Background and Research Questions}\label{sec:background}



\begin{figure}[ht]
    \centering
    \includegraphics[width=\columnwidth,keepaspectratio]{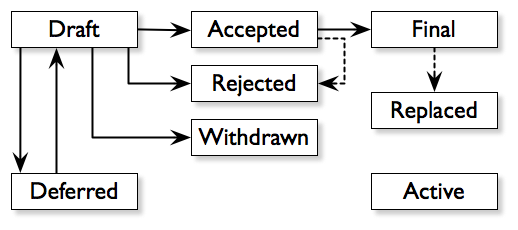}
	\caption{Decision-making process used in Python Enhancement Proposals~\cite{keertipati2016exploring}.}
	\label{fig:miningadditionalmainstates}
\end{figure}

Only recently have researchers explored the `actual' decision-making processes that lie hidden in OSS  repositories. The first focus had been the extraction of high-level decision-making processes in the form of a process diagram. Prior studies (\cite{savarimuthu2016process} and \cite{keertipati2016exploring}) extracted the decision-making process model using structured fields within messages from the \textit{Python-checkins} mailing list. The process extracted by study in \cite{keertipati2016exploring} 
showed 13 main states that a Python Enhancement Proposal (PEP) goes through. PEPs are documents that capture all the major proposed changes to the language, and also the processes that Python developers should adhere to. Whenever Python community members want to enhance the language with a new idea, feature, or patch, they propose it via a PEP. These proposals remain in a particular state (e.g., \emph{draft}) until a decision is made to move it to another state (e.g., \emph{accepted}). The official Python documentation showed eight decision making states for the PEP process, presented in PEP 1, as shown in Figure~\ref{fig:miningadditionalmainstates}. Based on mining Python repositories, prior work obtained five more decision-making states \cite{savarimuthu2016process} in the Python DM process. Subsequently, some researchers recently extended this study by extracting sub-states between the 13 main states \cite{sharma2020mining}. Their work revealed a richer and more complex decision-making process that encompassed sub-structures such as voting and consensus formation phases. However, a limitation of these prior works is that they did not capture the rationale behind state transitions. In other words, they did not answer the question {\em what is the rationale for a transition between states?} For example, how did a PEP move from {\em draft} to {\em accepted} state while another moved from \emph{draft} to {\em rejected}? Was it through \emph{consensus} or \emph{BDFL decree}? 

As previously noted, the rationale behind state changes in OSSD communities are often hidden inside email repositories. Identifying and extracting rationale from email messages is a challenging task for two reasons. First, mailing lists are informal communication channels and the rationale for state changes are thus often discussed in an informal way. There are no prescriptions or structure around how decisions should be communicated. Thus, rationale must be extracted from ambiguous and incomplete messages, which is a challenging computational task. Even a researcher armed with human intuition would need to manually analyze all the messages for a given PEP, and read the important messages several times, in order to understand the underlying rationale for a decision. Second, the underlying rationale for how decisions are made can be spread across many different email messages. These may need to be identified and linked in order to extract rationale.
	
This paper aims to overcome these challenges. Toward this goal, this work poses three research questions (RQs).

\begin{description}
	\item[\textbf{RQ~1}] \textbf{What are the different rationales for PEP decisions and their prevalence, as evident in Python email archives?}
	\item[\textbf{RQ~2}] \textbf{How can we design and develop an automated approach that extracts the rationale for PEP decisions?} 
	\item[\textbf{RQ~3}] \textbf{How effective is our approach at extracting the rationales behind PEP decisions?}
\end{description}

\section{Methodology}\label{sec:methodology}

\begin{figure}[ht]	
	\centering
    \includegraphics[width=\columnwidth,keepaspectratio]{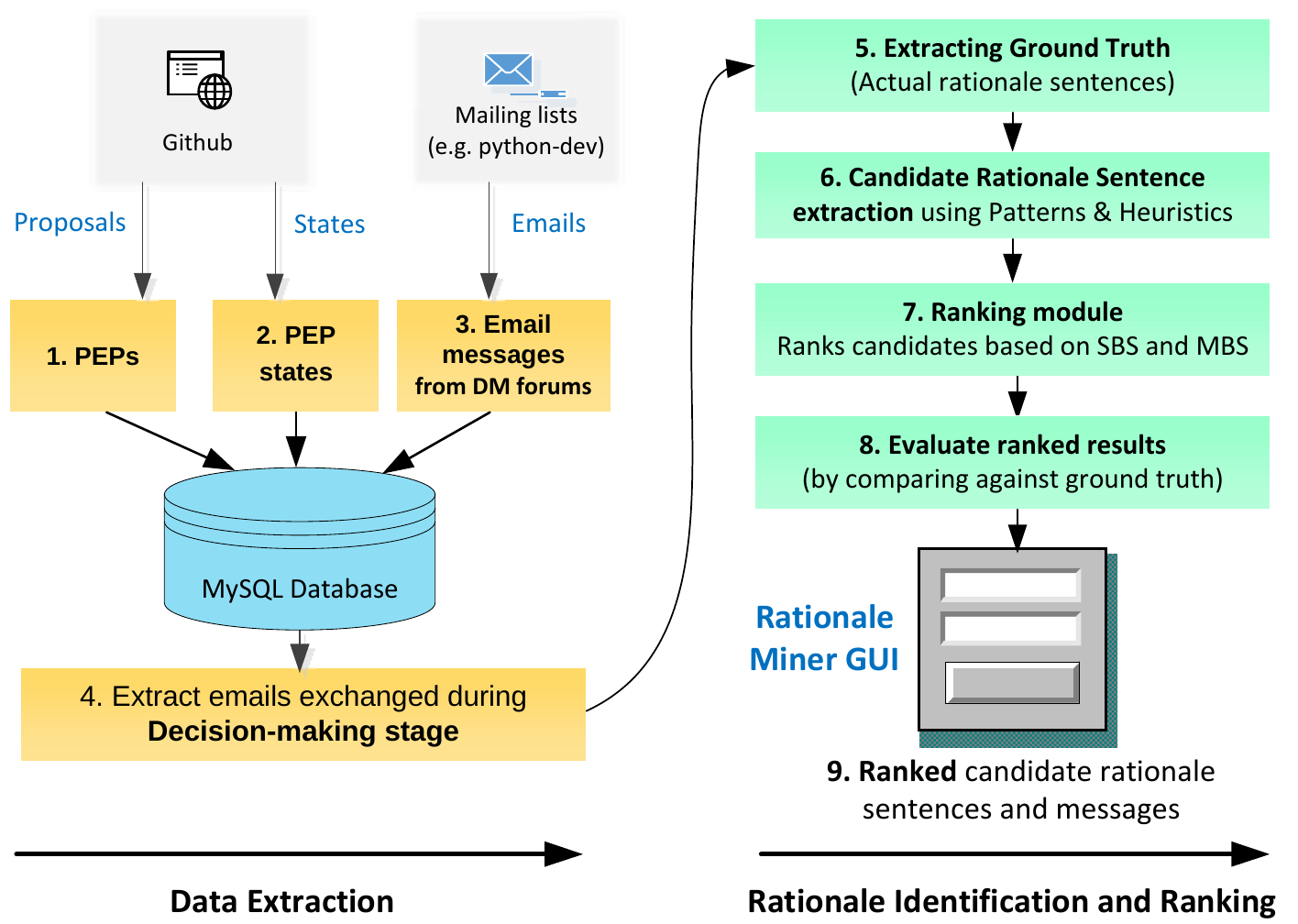}
	\caption{Methodology used to extract rationale behind PEP decisions.}
	\label{fig:heuristicreasonmodule}
\end{figure}

With regards to RQs 1-3, Figure~\ref{fig:heuristicreasonmodule} depicts the approach that we used to extract and present the rationale for decisions made during Python's evolution. This is embodied in a heuristics-based rationale extraction system called \emph{Rationale Miner}. Our approach comprises nine steps as outlined below.


    \textbf{Step 1:} The first step was to download all the 248 PEPs which were \emph{accepted} or \emph{rejected}. 
    Among other details, the PEP includes its author, and BDFL Delegates (if any) who made the final decision on the PEP. We obtained the PEPs from Python's Github repository\footnote{e.g., PEP 8: https://github.com/python/peps/blob/master/pep-0008.txt} and stored it in a 
    database.

    \textbf{Step 2:} Next, we retrieved all the 13 main states that each PEP transitioned through \cite{keertipati2016exploring}. This information was extracted for all PEPs from Python's GitHub repository, which includes all versions of every PEP document. These documents contain two key details: i) the next transition state of a PEP from its current state (e.g., from \emph{draft} to \emph{accepted}), and ii) the date of the change. We extracted Github data for Python state commits from its inception (March 1995) through to 12th July 2018. This end date was chosen because that was the date when the BDFL resigned, marking the end of benevolent dictatorship governance model. 

    \textbf{Step 3:} We then extracted all individual email messages from the Python email archives, and stored them in a MySQL database. We used the following email archives: \emph{Python-Dev}, \emph{Python-Ideas}, \emph{Python-Commits}, \emph{Python-Checkins}, \emph{Python-List}, and \emph{Python-Patches}.  These six mailing lists were chosen as they are the leading forums for Python developer discussions, with \emph{Python-Dev} being acknowledged as the main forum to discuss PEP development. The resulting dataset contained a total of 1,553,564 email messages. During this step, we also assigned a PEP number to email messages where possible. This was achieved by running a procedure that identified mentions of the PEP number(s) or title(s) within a message, and stored this as a field in the database alongside other details of a message (author, date, etc.).

    \textbf{Step 4:} From all the email messages stored in the database, we selected only those that related to PEP decision-making, i.e., emails that had successfully been assigned a PEP number.

   	{\textbf{Step 5:} \label{step:groundtruth} Next, we extracted the ground truth, i.e. the actual rationale behind PEP decisions, from the email messages of each PEP. This ground truth will be used to answer RQ1, and also to identify patterns of how rationale are stated while designing the automated rationale extraction approach. We also use this result to evaluate against the results of the Rationale Miner tool. Our approach in compiling the ground truth is explained in Section~\ref{sec:groundtruth}.

    \textbf{Step 6:} In this step, we identified a set of heuristics that could be used to separate rationale-containing sentences from non rationale-containing sentences. Scores were assigned to individual heuristics, and each sentence was assigned an aggregate heuristic score based on how well it matches the  with relevant heuristics. These heuristics and the relevant operational details to extract candidate rationale-containing sentences are described in detail in Section~\ref{sec:heuristicsbaseddss}.
    

    \textbf{Step 7:} The goal of this step was to rank candidate rationale-containing sentences using two schemes: a Sentence Based Scheme (SBS) and a Message Based Scheme (MBS). In SBS, candidate rationale-containing sentences from the previous step were ranked based on their heuristic score. In MBS, candidate sentences were aggregated at the message level and the ranks of the messages were computed. The ranks indicate the position of a sentence or a message respectively in the rank-ordered output, with the top~ranks 
    having 
    the highest heuristic score for containing rationale. Sections~\ref{sec:sentencelevelheuristics} and~\ref{sec:messagebasedscheme} explain these two approaches respectively.

    \textbf{Step 8:} Next, we evaluated the performance of the SBS and MBS by comparing their rank-ordered results with the ground truth (step 5). This is discussed further in Section \ref{sec:methodology_of_evaluation}. 

    \textbf{Step 9:} The final step was to present the rank-ordered results for both SBS and MBS in a graphical user interface (GUI). The GUI can be used to browse the ranked results to examine and understand the rationale behind a particular decision. This is discussed further in Section \ref{sec:GUI_description}. Subsequently we also obtained feedback from a former Python steering council member who was closely involved in decisions made in Python which are presented in Section \ref{sec:results}.

\subsection{Extracting Ground Truth} \label{sec:groundtruth}

To achieve our goal of automatically extracting rationale sentences we had to create a ground truth dataset comprising rationale for decisions that were made. To obtain the ground truth, we conducted an in-depth manual exploratory analysis of email messages (spanning multiple mailing lists) relating to all PEPs that reached either the \textit{accepted} (152 PEPs) or \textit{rejected} (96 PEPs) states. The first author read the corresponding messages and identified the rationale behind why each PEP was accepted or rejected. A custom-built GUI tool\footnote{A screenshot of the tool used to analyze messages can be viewed at https://doi.org/10.6084/m9.figshare.12363014.} was used for this purpose, which retrieved all messages for a PEP in date order. We examined messages for each PEP, paying particular attention to their decision date, in order to understand how the discussions unfolded and how the final decisions were made. When rationale for decisions were found (e.g., decision made based on consensus amongst developers) the rationale-containing sentences and the messages containing them were marked as ground truth data. 




Of the 152 \textit{accepted} PEPs, 107 had an explicitly stated rationale for acceptance. We manually extracted 179 rationale sentences from 162 unique messages relating to these PEPs. For some PEPs we found multiple rationale sentences. Of the 96 \textit{rejected} PEPs, 86 had an explicitly stated rationale for rejection. We manually extracted 121 rationale sentences from 97 unique messages relating to these PEPs.
This gave us a ground truth dataset of 300 rationale sentences relating to 193 PEPs. All rationale sentences in the dataset were verified by the second author (i.e., 100\% consensus).



The next section describes how we automated the process of rationale extraction, as manually extracting rationale from large datasets is laborious and time-consuming. The first author spent three months full-time on the exploratory analysis described in this section. This required reading most messages relating to the 248 PEPs, with some messages requiring multiple readings to fully understand the nuances.

\subsection{Heuristics-Based Approach to Extract Rationale}\label{sec:heuristicsbaseddss}

We used the heuristics-based approach to address the complexities of identifying rationale containing sentences. In the literature on software design decisions rationale extraction (\cite{williams2018using, teufel1997sentence, edmundson1969new, okurowski1999b, kurtanovic2017mining}), a number of heuristics have been proposed to be used in different contexts. For example, the work in \cite{kurtanovic2017mining} identifies users' rationale for writing app reviews. The work of Williams et al. \cite{williams2018using} have employed heuristics to extract arguments in software engineering practitioners' blogs. However, these works do not identify and extract the rationale behind \textit{`how'} these decisions are made in the OSS development context using a data-driven approach. 

Our approach, developed as the \textit{Rationale Miner} tool, uses 13 heuristics that were indicative of rationale on how PEP decisions are made, grouped into five categories. These heuristics were identified mainly from the literature and supplemented by our personal experience with manual rationale extraction from PEP messages. These heuristics were used to create a scoring function that is used in both the SBS and MBS. 

\subsubsection{Sentence-Based Scheme (SBS)}\label{sec:sentencelevelheuristics}

SBS considers five categories of heuristics to compute the heuristic score of a sentence.

\textbf{Category 1: Term patterns.} This category, comprising three heuristics, considers whether a sentence contains certain patterns of terms\footnote{\label{note:terms}The full list of rationale term types, their patterns and scores, see https://doi.org/10.6084/m9.figshare.12385862.} that indicate the presence of a rationale 
behind a PEP decision
\cite{python2003a}: ``Accordingly, the PEP \textit{[Proposal Identifier]} was rejected \textit{[State]} due to the lack of an overwhelming majority \textit{[Reason]} for change.'' Here, the term pattern is made up of three term types (in brackets) and the rationale for the decision (lack of majority) can be clearly inferred. These term patterns can appear in:
\begin{itemize}
    \item the sentence currently being considered, represented by the heuristic \emph{Term Pattern in Current Sentence (TPCS)};
    \item the remainder of the paragraph following the current sentence, represented by the heuristic \emph{Term Pattern in Rest of Paragraph (TPROP)}; and
    \item the subject line of the message containing the current sentence, represented by the heuristic \emph{Term Pattern in Message Subject (TPMS)}.
\end{itemize}
These three heuristics play an important role in determining whether a sentence may contain rationale. The scores assigned to the different term patterns range between 0 and 0.9 (see footnote 3).

Three of the six terms listed in the link in the footnote above were inspired from the literature on rationale identification. These are \textit{Reason Identifier}, \textit{Entity}, and \textit{Decision Terms} from the works of (\cite{li2020automatic}\cite{williams2019finding}\cite{kurtanovic2017mining}). 



\textbf{Category 2: Proximity-based heuristics.} We have also identified heuristics that are based on proximity. These include the location of the email message in relation to the PEP's evolution and the location of a sentence within a message. This category also comprises three heuristics.

\textit{Days From State Commit (DFSC):} Based on our manual analysis we found that messages containing rationale tended to appear closer to the dates of making decisions. When a decision is made (e.g., a PEP going from \emph{draft} to \emph{accepted}), these state changes are reflected in the PEP document committed to the version control system. Messages for a PEP that are closer (number of days) to the date of the commit recording the PEP's acceptance or rejection are assigned a higher score. 

\textit{Sentence Location In Message (SLIM):} Research has shown that the first sentence in a paragraph always introduces the main topics of the paragraph, while the last sentence in many instances summarizes what the paragraph discusses~\cite{edmundson1969new, mcdonald2002using, kupiec1999trainable, goldstein1999summarizing,yeh2005text, hu2017opinion}. We found evidence of this in our manual analysis of emails, where sentences located in the first or second paragraph conveyed the main idea of the message while the last paragraph placed emphasis on what had already been said. Therefore, in our heuristics scheme, if a sentence is in the first or last paragraph of a message, that sentence is assigned a score of 0.9 for this heuristic, otherwise 0.

\textit{Negation Terms (NT):} As PEP generally undergoes several rounds discussion, negation terms such as ``may'', ``not'', and ``should not'' are generally used in earlier stages of the PEP, not the final PEP decision. They change the meaning of a sentence. Therefore, if a sentence contains such terms, it is assigned a \(-\)0.8 negation penalty that reduces its total score. 

\textbf{Category 3: Role-based heuristics.}  A number of fields in an email message can play a role in identifying a rationale sentence. We group them into three types, each represented by a corresponding heuristic.
 
 \textit{Message Type (MT):} From our findings there are three types of email messages related to a PEP: the official PEP summaries, PEP state commit messages, and all other messages pertaining to the PEP. The \textit{PEP summaries} are official summaries that reflect on the PEP, written after a decision has been made and placed on the Python website.\footnote{e.g., for PEP~308: https://www.python.org/dev/peps/pep-0308.} 
 \textit{State commit} messages reflect state changes (e.g., \emph{draft} to \emph{accepted}), as previously mentioned. If a sentence belongs to one of these two message types, it has a higher chance of containing a rationale, and is therefore assigned a score of 0.9; otherwise it scores 0.
 
\textit{Author Role (AR):} During the different stages of a PEP's evolution, various members of the Python community write different types of email messages for a PEP. However, messages containing decisions and their rationale are usually written by certain roles. If the message is from the \textit{BDFL} or \textit{BDFL Delegate} (who may differ for each PEP), the sentences in the message are assigned a score of 0.9 for this heuristic. If the message is from a \textit{PEP Author} then this heuristic is assigned a score of 0.6. If the message is from a \textit{PEP Editor} (who are mainly consulted in the earlier stages of the PEP evolution) the score assigned is 0.5, while messages from a \textit{Core Developer} score 0.4. A score of 0 is assigned to this heuristic if the message is written by anyone else (i.e. developers or users). This heuristic is inspired based on prior works (\cite{okurowski1999b}\cite{mcdonald2002using}\cite{hu2017opinion}\cite{ye2003toward}) that have considered the roles played by individuals (e.g. author of a message).

\textit{Specific Author Messages Containing Explicit Rationale (SAMCER):} This heuristic identifies messages that contain specific hints about rationale.

\begin{enumerate}
    \item \textit{PEP author's message containing rationale:} When a PEP author feels the PEP is nearing completion, they write a message formally requesting a review and pronouncement on the PEP. The community's consensus on the PEP is sometimes mentioned in this message, for example: ``As you said, consensus is reached, so just Guido's BDFL stamp of approval is all I can think of''~\cite{python2010a}. Here, consensus has been reached in the community and the PEP is waiting for a pronouncement from the BDFL. The rationale for the PEP's eventual approval is the indicated community consensus.
	\item \textit{BDFL or BDFL Delegate PEP Review:} In messages preceding formal acceptance of the PEP, the BDFL or delegate may mention their rationale for future acceptance. For example, ``Assuming no material objection [\emph{sic}] appear to the new syntax and semantics, I can approve the PEP later this week'' \cite{python2015a}. This PEP was accepted few days later, the rationale being that the community had no further objections (i.e., lazy consensus).
	\item \textit{BDFL or BDFL Delegate PEP Acceptance or Rejection:} The BDFL or delegate formally accepts or rejects the PEP, but the rationale for acceptance is not explicitly mentioned.  For example: ``Given the feedback so far, I am happy to pronounce PEP 393 as accepted'' \cite{python2011a}. In this case, the PEP was accepted because of community consensus, which is implicit in the sentence.
	\item \textit{Community members reflecting on decisions:} Sometimes a core developer would mention how a PEP was received by the community, either while summarizing the PEP or discussing the decision later on. For example: ``Raymond suggested updating PEP 284 (`Integer for-loops'), but several people spoke up against it, including Guido, so the PEP was rejected'' \cite{python2005b}.
\end{enumerate}

If a sentence belongs to a message from one of these four types, it is assigned a score of 0.9, otherwise 0.

\textbf{Category 4: Response to certain specific messages.} This category comprises two heuristics related to 
specific messages.

\textit{Response Messages to the Same State Change Message (RMSSCM):} If the PEP author requests pronouncement based on feedback received from the community, for example by stating ``there were no outstanding issues'', it is likely that the reply message will contain the rationale for the subsequent decision. Therefore, we assign a higher score to messages with the same subject line. This heuristic was inspired by a similar heuristic used by AbdelRahman et al.\ \cite{abdelrahman2010new}. 

\textit{Rationale Found Using Triple Extraction (RFUTE):} In our previous research work (\cite{sharma2020mining}), we used Subject, Verb, Object (S,V,O) triples to identify decision-making sub-states that occur between the main states shown in Figure~\ref{fig:miningadditionalmainstates}. In our manual analysis, we observed that we could automatically identify some sentences as rationale-containing sentences via the S,V,O triples they contained. For example, the S,V,O triple extracted from the rationale sentence ``As you said, consensus is reached, so just Guido's BDFL stamp of approval is all I can think of''~\cite{python2010a} is [``consensus'', ``is'', ``reached''] which implies consensus. We therefore adopted this approach to identify and match S,V,O triples, where certain decision-specific terms, such as ``consensus'', were included in the triples. Sentences where such S,V,O triples were extracted and matched were assigned a score of 0.9, otherwise 0. 

 

\textbf{Category 5: Special Identifiers.} Based on our analysis of PEP email messages, we included two additional heuristics for special identifiers that exhibited rationale behind state changes. These heuristics have not been previously reported in the literature on rationale identification.

\textit{Decision Terms In Message (DTIM):} On some occasions, members inserted a heading before the paragraph that stated the rationale for PEP acceptance or rejection. Examples include ``BDFL Pronouncement'', ``PEP Acceptance'', and ``PEP Rejection''. We consider these as terms that convey decisions and therefore any sentence containing such terms is assigned a score of 0.9 for this heuristic, otherwise, 0.

\textit{Decision Terms as Header of current Paragraph (DTHP):} If the decision terms identified above exist as the heading of the paragraph that contains the sentence currently being considered, we assign a score of 0.9 for this heuristic to all sentences in that paragraph.

\textbf{The Final Scoring function.} The Final Score (\(\mathit{FS}\)) for each sentence is the sum of its scores for all 13 heuristics:
\begin{multline}\label{eq:scoringfn2}
\mathit{FS} = \mathit{TPCS} + \mathit{TPROP} + \mathit{TPMS} + \mathit{DFSC} \\
    + \mathit{SLIM} + \mathit{NT} + \mathit{MT} + \mathit{AR} + \mathit{SAMCER} \\
    + \mathit{RMSSCM} + \mathit{RFUTE} + \mathit{DTIM} +  \mathit{DTHP}
\end{multline}
The sentence scores of all sentences belonging to a PEP are compared in order to produce a descending rank-ordered list of candidate rationale containing sentences.

\subsubsection{Message-Based Scheme (MBS)}\label{sec:messagebasedscheme}

In the sentence-based scheme, candidate rationale sentences from the same email message may appear in different rank-ordered positions, as the rationale may be present in multiple sentences. However, a user may prefer to see results from the perspective of the entire message containing these rationale sentences.
MBS is a message-level aggregation of the SBS. This approach has two advantages. First, it prevents multiple rationale sentences from the same message being shown in different rank-ordered positions, so that users can view related rationale sentences in the context of the same message. This will reduce the effort required for the user to scroll through the list of rationale sentences. Second, and more importantly, it shows the entire message which may provide a richer context around the rationale for the candidate rationale sentence. The evaluation of these two approaches is presented in Section~\ref{sec:comparison}.

\subsection{Heuristic optimization based on ground truth}  \label{sec:varsel}



After identifying the 13 heuristics described above, we undertook two-pronged optimization: 1) identifying the heuristics (variables) that strongly influence the identification of rationale and 2) using parameter sweeping to find the best values for these variables. We describe these two aspects next.

Having set values for the 13 heuristics to initial values as described in the previous section, we computed the scores for each sentence in all the emails. Then, we computed the total number of sentences that were correctly identified in the top-5 results in the SBS scheme (by checking whether a rationale containing sentence (from the ground truth set) appeared in these results. Having computed this result (which we call the baseline result for optimization purposes), we systematically removed one variable at a time and compared how that impacted the overall results. The goal of this approach was to identify variables that strongly influence our results. 

\textbf{Identifying influential heuristics} - Table \ref{tab:relativeimportancebaselineresults} shows the number of fewer or additional rationale sentences captured at top-5 ranks for \textit{Accepted} and \textit{Rejected} PEPs in both the SBS and the MBS after removing each heuristic from the $FS$. The negative values indicate the removal of the associated heuristic 
decreases the number of rationale sentences captured: thus it has a positive influence. The influence of the variables are divided into five groups as shown in Table \ref{tab:relativeimportancebaselineresults}.

\begin{table}[htb]
	\centering
	\caption[Relative contribution of heuristics in top 5 rankings in the baseline results]
	{Relative contribution of heuristics in top 5 rankings 
	}
	\label{tab:relativeimportancebaselineresults}
	\begin{tabular}{lccccc}
		\toprule
		& \multicolumn{2}{c}{\textbf{SBS}} 
		& \multicolumn{2}{c}{\textbf{MBS}}
		& \multirow{2}{*}{\textbf{Influence}} \\ 
		\cmidrule(r){2-3} 
		\cmidrule(l){4-5} 
		\textbf{Heuristics} & \textit{Acc.} & \textit{Rej.} & \textit{Acc.} & \textit{Rej.} &  \\ 
		\midrule
		RFUTE & -12 & -5 & -14 & -9 & \multirow{2}{*}{\begin{tabular}[c]{@{}c@{}}Positive\\ (strong)\end{tabular}} \\ 
		TPCS & -5 & -3 & -4 & -2 &  \\ 
		\midrule	
		DFSC & -3 & -1 & -6 & -6 &  \multirow{2}{*}{\begin{tabular}[c]{@{}c@{}}Positive\\(medium)\\ \end{tabular}} \\ 
		DTIM  & -1 & 0 & -1 & -2 & \\ 
		\midrule
		
		SAMCER & 4 & 2 & -1 & 0 & \multirow{3}{*}{\begin{tabular}[c]{@{}c@{}}Mixed\\ (strong)\end{tabular}} \\ 
		MT & 5 & -2 & -2 & -1 &  \\ 
		{\begin{tabular}[c]{@{}c@{}}TPMS  \end{tabular}}
		& 7 & -2 & -8 & -1 &  \\ 
		\midrule
		DTHP & 1 & 0 & -1 & -1 & \multirow{3}{*}{\begin{tabular}[c]{@{}c@{}}Mixed\\(weak)\\ \end{tabular}} \\ 
		NTP & 1 & 0 & 0 & -1 &  \\ 
		{\begin{tabular}[c]{@{}c@{}}TPROP \end{tabular}} 
		& 2 & -1 & -1 & -3 &  \\ 
		\midrule
		SLIM & 0 & -2 & 0 & -1 & \multirow{3}{*}{None} \\ 
		AR & 0 & -1 & 0 & 0 &  \\ 
		RMSSCM & 0 & 0 & 0 & -1 &  \\ 
		\bottomrule
	\end{tabular} 
\end{table}

The \textit{positive influential heuristics} were RFUTE and TPCS (both strong). When these variables were removed, the results were negatively impacted (e.g., up to 12 results not appearing in the top-5 results). DFSC and DTIM variable had a medium (positive) influence on the results. 
The heuristics that had \textit{mixed but strong influence} were SAMCER, MT and TPMS. Their influences on both types of PEPs and in both schemes were \textit{strong}, but varied (i.e., values are positive at times and negative at other times), and thus were of particular interest during optimisation (see next paragraph) 
Heuristics with \textit{weak mixed influence} or those with almost \textit{no} effect the top 5 rankings were DTHP, NTP, TPROP, SLIM, AR, and RMSSCM. 



\textbf{Parameter Sweeping:} 
For the seven variables that either had positive influence or mixed influence, we performed parameter sweeping, a technique widely used in simulation systems \cite{buyya2005scheduling} to identify the best values for the variables by changing the values in increments of 0.3. Five values were considered for each heuristic (-0.3, 0, +0.3, +0.6, +0.9). We varied one heuristic at a time by keeping all the values of all the other heuristics the same. This approach enabled us to find the best configuration of values for the chosen heuristics, by comparing the results obtained by a configuration with the ground truth results. 
Adopting this approach, we found that for both SBS and MBS, generally, our current DFSC and TPMS heuristics values needed to be incremented by 0.6. Changing the values for the other heuristics made no significant difference. Hence, the results reported in Section \ref{sec:results} are based on the best configuration for these 13 heuristics.


\subsection{Evaluating the Heuristics-Based Approach}	\label{sec:methodology_of_evaluation}


We compared the results obtained using the heuristics-based approach against the ground truth using two methods. First, we compared the rank-ordering of sentences and messages. If a sentence was identified in the ground truth dataset as a rationale-containing sentence for a particular PEP, we checked the rank of the same sentence (and its corresponding message) in the heuristics-based results. If there was only one rationale sentence for a PEP in the ground truth dataset, then the ideal result would be that the heuristics-based approach assigned the highest score (i.e., highest rank) to that sentence and its corresponding message. We,  therefore, quantified the number of sentences and messages that matched at various ranks.

Second, we used the \emph{normalized discounted cumulative gain (NDCG)} metric, a metric commonly used in the information retrieval domain to evaluate the ranking results for a query in relation to the ``perfect'' ranking for the same query~\cite{croft2015search}. It has, for example, been used to evaluate an email ranking system based on search query results~\cite{abdelrahman2010new}. In the context of our work, most PEPs have one or two rationale sentences. If for instance these are not captured in the top two ranks, the ranking quality for these two positions will be penalized by this approach. NDCG is calculated using this formula~\cite{croft2015search}:
\begin{equation} \label{eqn:NDCGatk} 
    \mathit{NDCG}_{k} = \frac{\mathit{DCG}_{k}} {\mathit{IDCG}_{k}}
\end{equation}
where \(\mathit{IDCG}_{k}\) is the ideal DCG value (based on its definition in \cite{croft2015search}) of the ranking at position \(k\), that is, the DCG value of the ground truth. NDCG is normalized to the interval [0,1], with 1 indicating a perfect estimation of the ground truth.

\section{Results}\label{sec:results}

This section presents the results for the research questions. 

\subsection{RQ1: Rationales for PEP Decisions}

Figure~\ref{fig:dmconceptspiechart} shows 11 different decision rationales for \textit{accepted} and \textit{rejected} PEPs that have been unearthed in our qualitative analysis. 
These are based on the 300 ground truth rationale sentences we manually extracted.\footnote{A spreadsheet containing the raw data for these sentences can be viewed at https://doi.org/10.6084/m9.figshare.12732260.} Table~\ref{tab:reasonssentencetable} gives examples of rationale sentences corresponding to each of these rationale.

\begin{figure}[htb]
	\centering
	\includegraphics[width=\columnwidth,keepaspectratio]{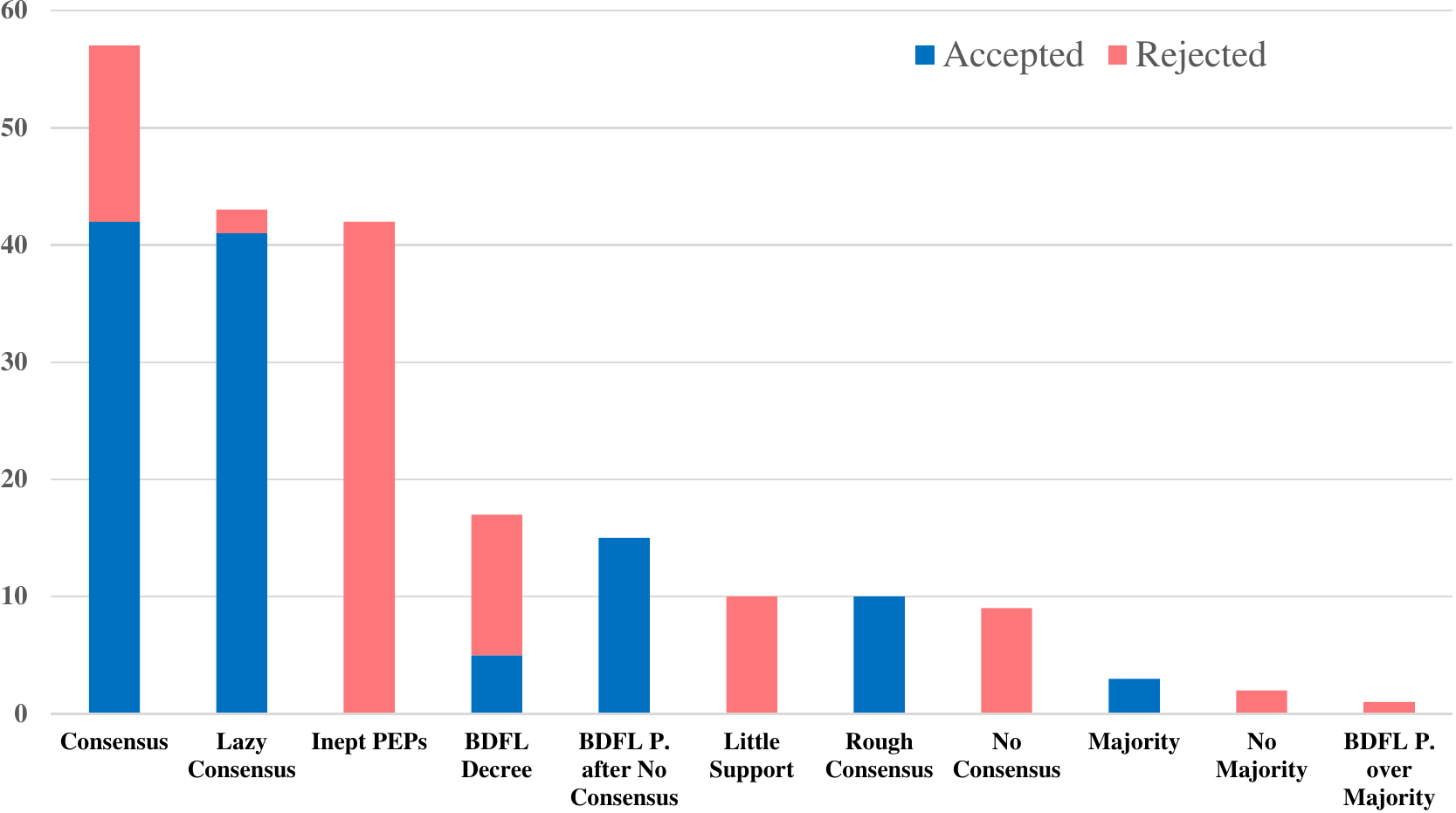}
	\caption{Rationale for \textit{accepted} and \textit{rejected} PEPs. The horizontal axis shows the rationale and the vertical axis the number of PEPs.}
	\label{fig:dmconceptspiechart}
\end{figure}

\begin{table*}[ht]
	\centering
	\caption{Sample sentences containing rationale behind decisions on \textit{Accepted} and \textit{Rejected} PEPs}
	\label{tab:reasonssentencetable}
	\begin{tabular}{llr}
		\toprule
		
		\textbf{Rationale} & \textbf{Rationale sentence} & \textbf{PEP}  \\ 
		\midrule
		Consensus & ``The user community unanimously rejected this so I won't pursue this idea any further'' \cite{python2001c} & 259 \\
		No Consensus & ``Although a number of people favored the proposal there were also some objections
		'' \cite{python2007b} & 3128 \\
		Lazy Consensus & ``If anyone has objections to Michael Hudson's PEP 264: raise them now.'' \cite{python2001b} & 264 \\
		Rough Consensus & ``Several people agreed, and no one disagreed, so the PEP is now rejcted. [\emph{sic}]'' \cite{python2005b} & 265 \\
		Little Support & ``It has failed to generate sufficient community support in the six years since its proposal.'' \cite{python2001a} & 268 \\
		Majority & ``Comments from the Community:  The response has been mostly favorable''. \cite{python2002c} & 279 \\
		No Majority & ``Accordingly, the PEP was rejected due to the lack of an overwhelming majority for change.''\cite{python2003a} & 308 \\
		BDFL Decree & ``After a long discussion, I've decided to add a shortcut conditional expression to Python 2.5.''  \cite{python2005a} & 308 \\
		BDFL P. after No Consensus	& ``There's no clear preference either way here so I'll break the tie by pronouncing false and true.'' \cite{python2002a} & 285 \\		
		BDFL P. over Majority & ``Python is not a democracy.'' \cite{python2004a} & 326 \\
		\bottomrule
		\multicolumn{3}{l}{Most of these statements were made by the BDFL directly or are references to the BDFL's views.}
	\end{tabular}	
\end{table*}


The top~5 rationale for decisions were \textit{consensus} (57), \textit{lazy consensus} (43), \textit{inept PEPs} (42), \textit{BDFL decree} (17) and \textit{BDFL pronouncement after no consensus} amongst developers (15). \emph{Inept PEPs} is an aggregate of eight rationale that represent a problem with a PEP, such as ``PEP benefits marginal'', ``requires significant changes'', etc. \emph{BDFL decree} refers to instances where the BDFL makes a decision regardless of the community's views. The descriptions of each of the rationale is provided in this link\footnote{https://doi.org/10.6084/m9.figshare.12887885}. 

The work in \cite{mrowka2012decision} outlines the commonly used rationale behind decisions (i.e. `how' decisions are made)  by considering the Apache OSS project. The work identifies five rationales specific to this project. 
However, no prior work has been undertaken for Python to understand the rationale in decision-making in the context of PEPs, and in particular conducting a data-driven qualitative study to unearth the `actual'  rationale from email discussions and the prevalence of the usage of each rationale as we have undertaken here. We have also unearthed several additional rationales (e.g., \textit{rough consensus} and the ones involving BDFL) whose prevalence has not been previously quantified in the literature. To our understanding, \textit{rough consensus} is used to consider some form of community majority when full consensus is not attainable. Some PEPs have intense discussions where it is not possible to achieve full consensus, and \emph{rough consensus} may be a way to avoid a situation where the community's preference is based only on the opinions of a vocal minority or influential individuals~\cite{mailarchive2009a}.

There are rationale whose occurrences indicate a clear decision (e.g., \emph{accepted} or \emph{rejected}, but not both). For example, \textit{little support}, \emph{no consensus}, \emph{no majority}, and \emph{lack of champion} always lead to a PEP being \emph{rejected}. There are other rationale, however, where the final decision is not so clear cut. For example, even after \emph{lazy consensus} or \emph{rough consensus}, a PEP could be still rejected (possibly due
to \emph{BDFL decree}). We also observed one PEP (326) where the BDFL overrode the majority's opinion. The majority of the community preferred the PEP, but the BDFL rejected it, emphasizing that ``Python is not a democracy'' \cite{python2004a}. The substantial number of PEPs where the BDFL has exercised his own preference (\textit{BDFL decree}, \textit{BDFL pronouncement after no consensus}, and \textit{DFL pronouncement after majority}) implies that in the Python community, the BDFL is free to exercise his choices when necessary. However, most of the decisions taken by the BDFL, are based entirely on a collective community view,  
such as \textit{consensus, lazy consensus, rough consensus, majority, no majority, and little support}. The Python project leader allows the community to come to a collective view on proposal outcomes, using any of these rationales, and then he (or sometimes his delegates) appear to mostly concur with the collective view.

\begin{figure*}[ht]
	\centering
	\includegraphics[width=\textwidth,height=8cm]{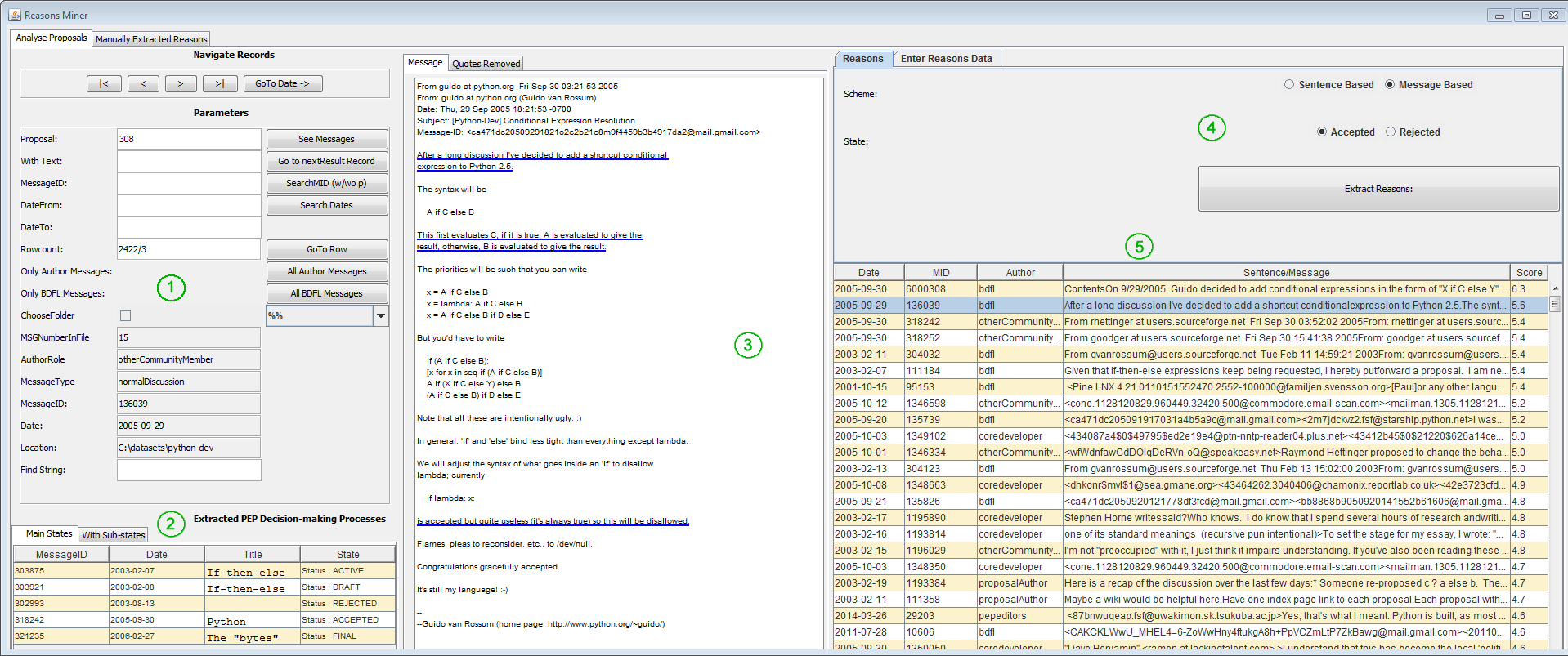}
	\caption{The Rationale Miner GUI. (A larger version may be viewed at https://doi.org/10.6084/m9.figshare.12377633)}
	\label{fig:reasonsminergui}
\end{figure*}

\subsection{RQ2: An Approach for Rationale Extraction}\label{sec:GUI_description}
To answer RQ2, we first identified relevant heuristics both from the literature and from our efforts during the manual extraction of rationale-containing sentences. We identified 13 such heuristics. Using these heuristics we then constructed a scoring mechanism that was used to rank candidate rationale-containing sentences in the SBS. At a higher level of aggregation, messages containing these rationale sentences were ranked using the MBS.

Figure~\ref{fig:reasonsminergui} shows the Graphical User Interface (GUI) of the \textit{Rationale Miner}. This is designed to help users understand the rationale behind how decisions made, by enabling them to explore the results of both sentence-based and message-based rationale extraction for a particular PEP. Panel~1 (green numbering) is for entering search parameters such as the desired PEP number. Panel~2 shows a timeline of states traversed by the selected PEP (e.g., \emph{draft} and \emph{accepted}). To obtain all the rationale sentences (or messages), the user can select in Panel~4 the appropriate approach (sentence-based or message-based) and also whether they are interested in PEPs that were \emph{accepted} or \emph{rejected}. Clicking the \emph{Extract Reasons} button causes the results to be shown in Panel~5. Clicking on a result row in Panel~5 shows the corresponding sentence or message in Panel~3. When a message is shown in Panel~3, all rationale-containing sentences are underlined. A video demonstrating the features of the GUI can be viewed at https://youtu.be/nrB9Jk1OFXo. For simplicity, in the video, the Rationale Miner is referred to as the \textit{Reasons Miner}.

\subsection{RQ3: Effectiveness of the Approach}\label{sec:comparison}
In this section, we demonstrate the effectiveness of the Rationale Miner in identifying rationale using two approaches.

\subsubsection{Evaluation Approach 1---Comparing Rank-Ordered Results}\label{sec:rank-ordered}
Table~\ref{tab:reasonsoptimisedmetricsforacceptedandrejected} shows the number and percentage of ground truth rationale-containing sentences captured by SBS and MBS at a particular rank \(k\), for PEPs that reached the final states \textit{accepted} or \textit{rejected}. We can see that 15.7\% (\emph{accepted}) and 29.8\% (\emph{rejected}) of the ground truth sentences appeared as the top result (\(k = 1\)) in the heuristics-based system using SBS. MBS performed even better: 20.2\% and 47.1\%, respectively. Furthermore, 39.7\% (\emph{accepted}) and 48\% (\emph{rejected}) of ground truth sentences appeared within the top~5 results using SBS (61.5\% and 74.4\% respectively for MBS, and highlighted in bold in Table~\ref{tab:reasonsoptimisedmetricsforacceptedandrejected}). The number of ground truth sentences appearing within the top \(k\) SBS and MBS results increases with \(k\), as to be expected, but it is heartening to see that a relatively large fraction of ground truth sentences are ranked within the top~5, especially for MBS. Most of the rationale sentences for both states are captured within the top~100.



\begin{table}[htb]
    \centering
    \caption{Heuristics-based approach vs.\ ground truth}
    \label{tab:reasonsoptimisedmetricsforacceptedandrejected}
    \begin{tabular} {lrrrrrrrr}
        \toprule
            & \multicolumn{4}{c}{\textbf{SBS}} &  \multicolumn{4}{c}{\textbf{MBS}} \\
        \cmidrule(r){2-5} 
        \cmidrule(l){6-9} 
            & \multicolumn{2}{c}{\emph{Accepted}} & \multicolumn{2}{c}{\emph{Rejected}} & \multicolumn{2}{c}{\emph{Accepted}} & \multicolumn{2}{c}{\emph{Rejected}} \\
        \cmidrule(r){2-3}
        \cmidrule(lr){4-5}
        \cmidrule(lr){6-7}
        \cmidrule(r){8-9}
        \smash{\shortstack[l]{\textbf{Ranking} \\ \textbf{(Top~\emph{k})}}} & CC\rlap{\textsuperscript{*}} & \% & CC & \%& CC & \%& CC & \% \\
        \midrule
        Top~1    & 28  & 15.7 & 36 & 29.8 & 36  & 20.2 & 57  & 47.1 \\
        Top~2    & 30  & 16.7 & 42 & 34.7 & 67  & 37.4 & 68  & 56.2 \\
        Top~3    & 45  & 25.1 & 49 & 40.5 & 85  & 47.5 & 79  & 65.3 \\
        Top~4    & 52  & 29.1 & 56 & 46.3 & 103 & 57.6 & 85  & 70.2 \\
        \textbf{Top	5} & \textbf{71} & \textbf{39.7} & \textbf{58} & \textbf{48.0} & \textbf{110} & \textbf{61.5} & \textbf{90} & \textbf{74.4} \\
        Top~10   & 94  & 52.5 & 72 & 59.5 & 133 & 74.3 & 104 & 86.0 \\
        Top~15   & 110 & 61.4 & 82 & 67.8 & 147 & 82.1 & 110 & 91.0 \\
        Top~30   & 131 & 73.2 & 94 & 77.7 & 161 & 90.0 & 111 & 91.8 \\
        Top~50   & 139 & 77.7 & 98 & 81.0 & 170 & 95.0 & 111 & 91.8 \\
        Top~100  & 158 & 88.3 & 99 & 81.8 & 175 & 97.8 & 112 & 92.6 \\
        Top~100+ & 0   & 0    & 0  & 0    & 0   & 0    & 0   & 0 \\
        No match & 21  & 11.7 & 22 & 18.2 & 4   & 2.2  & 9   & 7.4 \\
        \midrule		
        \textbf{Total} & 179 & & 121 & & 179 & & 121 & \\
        \bottomrule	
        \multicolumn{9}{l}{\textsuperscript{*}CC = cumulative count of rationale sentences} \\
    \end{tabular}	
\end{table}


Comparing the SBS and MBS results, two aspects can be noted. First, MBS captures a larger number of rationale sentences than SBS at each rank \(k\). This is because a ranked message in MBS can have more than one rationale-containing sentence. For example, suppose that there are two sentences from the same message ranked at 3 and 8 in the SBS ranking. Under MBS, the rank for this message will be 3, and both sentences will be considered part of this message. So if we consider the 9th ranked sentence under SBS, it will be ranked 8th under MBS. MBS thus has an inherent advantage over SBS. Second, the results show that there are more unmatched sentences under SBS than MBS. Unmatched sentences are those which were identified as rationale sentences in the ground truth dataset, but did not appear in results of the heuristics-based system. This is because these sentences did not have patterns matching the heuristics. MBS has fewer unmatched sentences because sentences not matched by SBS may appear in the same message as sentences that are matched.



We observe that many rationale are captured at lower ranks (say \(k >\textrm{15}\)). This mainly arises when rationale are stated long before or long after the date a PEP's \textit{accepted} or \textit{rejected} state was committed to the Python repository. For example, the rationale for a PEP's acceptance or rejection is conveyed in a message, but the PEP is only formally \textit{accepted} or \textit{rejected} much later (sometimes six months or in the extreme case up to two years later). Alternatively, the community member in charge of committing the decisive state of the PEP may for some reason
postpone doing so and commit the state changes for several outstanding PEPs as a batch. Since our approach prefers rationale sentences that are closer to the dates at which decisions were made, sentences written much earlier or much later are given a lower score in the DFSC heuristic, resulting in a lower rank.


\subsubsection{Evaluation Approach 2---Using NDCG}\label{sec:effectiveness}

We used the NDCG metric  to formally evaluate our heuristics-based approach, as discussed in Section~\ref{sec:methodology_of_evaluation}. We computed the average NDCG for each rank \(k\) (equation~\ref{eqn:NDCGatk}), for accepted and rejected PEPs under both SBS and MBS retrieval-based ranking  approaches are summarized in Figure \ref{fig:evaluationmetricsforHBDSS}. 
We used different thresholds for the minimal size of the search results, where the ranges over the values top 5; top 10; top 15; top 30; top 50; top 100.

\begin{figure}[htb]
	\centering
    \includegraphics[width=\columnwidth,keepaspectratio]{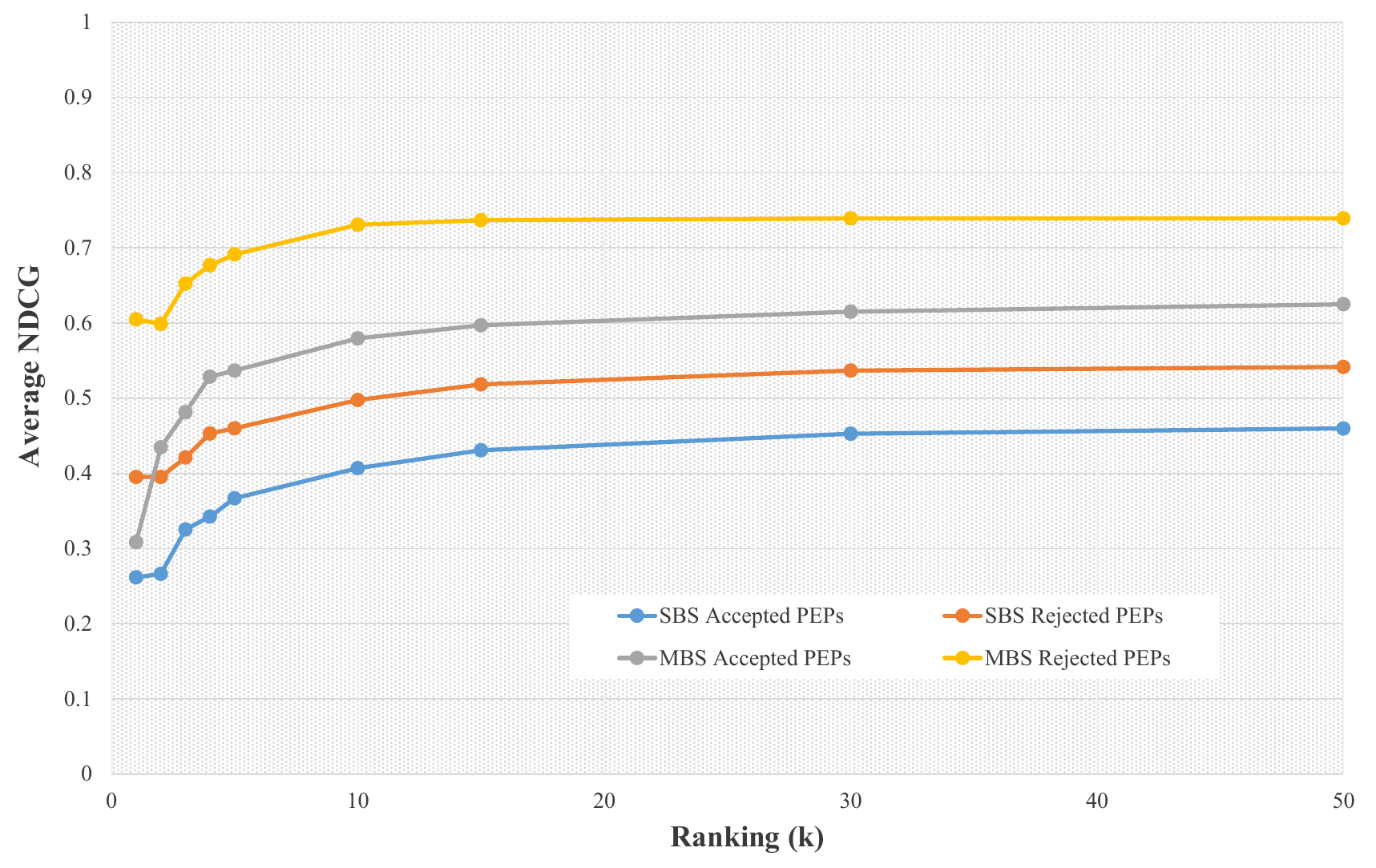}
	\caption{Average NDCG values of 152 \textit{Accepted} and 96 \textit{Rejected} PEPs for both MBS and SBS.}
	\label{fig:evaluationmetricsforHBDSS}
\end{figure}
 
The chart shows that MBS produces better results than SBS for both \textit{accepted} and \textit{rejected} PEPs at almost all ranks. This aligns with the results of the rank-ordering comparison in Section~\ref{sec:rank-ordered}. For example, the MBS scores for \emph{rejected} are higher than the SBS scores for all ranks \(k\). Specifically, MBS for \emph{rejected} outperforms SBS by 10.3\% at \(k = \textrm{5}\), 12.4\% at \(k = \textrm{10}\), 14.7\% at \(k = \textrm{25}\), and 11.2\% at \(k = \textrm{50}\).





\section{Discussion}\label{sec:discussion}

In this section we revisit our research questions and discuss the main contributions.


\subsection{RQ1: Rationale behind PEP decisions}

We can infer that while the Python OSS community employs a strict hierarchical and authoritative DM structure, the DM structure is not totally authoritarian. There is a large adherence to the community's view, acquired and represented via different mechanisms. For most PEPs, the project leader transfers the burden of coming to a decision to the community through some form of \emph{consensus} (including \emph{lazy consensus} and \emph{rough consensus}). This helps avoid the situation of the dictator's decision continually suppressing the views of the community. The BDFL uses this mixture of two principles, which seem to work well together, for the success of decision-making and the Python project overall. 

To evaluate the correctness of our results (i.e., the identified rationale are correct), we presented our findings to a prominent Python core developer (name withheld) who had been lead author or co-author of 22 PEPs and has been BDFL Delegate 14 times, and is a former Python Steering Council member. He replied \textit{``I think your list of reasons looks good''.} This demonstrates that the extracted rationale are accurate.


\subsection{RQ2: An Approach for Rationale Extraction}

We have operationalized a heuristic-based Rationale Miner tool to extract rationales. In doing so we automated the extraction of features in a sentence (e.g., presence of PEP number), the assignment of scores for a feature, and the final computation of overall score for a sentence that represents whether or not a sentence contains rationale or not. The two schemes for rationale identification the SBS and MBS offer distinct advantages. SBS offers a more fine-grained sentence-level rationale, while MBS offers coarse-grained message-level rationale. MBS can be argued to provide a richer context and thus a deeper understanding of the rationale behind decisions that were discussed on the mailing lists. However, SBS presents a more direct and explicit rationale-containing sentence that may be quicker and easier to understand. MBS presents entire messages instead of individual sentences, which will take more effort for users to read and understand. We have reduced this cognitive load in the Rationale Miner GUI (Figure~\ref{fig:reasonsminergui}) by highlighting the rationale-containing sentences within a message.

We asked the same Python core developer mentioned in the previous sub-section, whether the idea of a tool which extracts how PEP decisions are made, is useful. He emphasized the importance of such a tool in his response: \textit{``Yes, I think the idea is valuable, as it should be possible for at least Steering Council members to compare their subjective impressions of outcomes and rationales with the output of the tool, and potentially use that comparison to help assess how well the PEP process is actually working.''}

\subsection{RQ3: Effectiveness of the Approach}\label{sec:resultsdisevaluationmetrics}

The 
MBS is the best performing of the two schemes described in this work. It accurately (i.e., top-ranked result) captures 20\% of ground truth rationale sentences for \emph{accepted} PEPs and 47\% of sentences for \emph{rejected} PEPs (see Table~\ref{tab:reasonsoptimisedmetricsforacceptedandrejected}). If we consider the top~10 ranked results, MBS captures 74\% and 86\% of rationale sentences, respectively; and top~15 increases this further to 82\% and 91\%, respectively. Thus, we believe, the results are good enough for the intended purposes.

Having said the above, there are three main reasons why not all rationale-containing sentences are captured at the first rank. First, in some sentences, there may be little or nothing to indicate the presence of a rationale. For example, consider the sentence ``Augmented assignment is scheduled to go in soon (well, before 2.0b1 at least) and if you don't spot the loony now, we'll have to live with it forever :)''\cite{python2000a}. The implied rationale here is that there is {\em lazy consensus} on the PEP. This might be clear to a human, but there are no obvious patterns that can be identified here from the viewpoint of heuristics. 

A second reason is that some sentences do not include term patterns that commonly appear in rationale-containing sentences. For example, in PEP 3131, the sentence which comes closest to a rationale-containing sentence is: ``After 175 replies and counting the only thing that is clear is the controversy \emph{[Reason]} around this PEP \emph{[Proposal Identifier]}''\cite{python2007a}. There are identifiable two term types (in brackets) in this sentence, but these do not conform to the common patterns of how rationale are stated. These two term types are common in sentences, but not every sentence containing these term types will be a rationale sentence. Including the kinds of term patterns exhibited by the sentence above would therefore result in an unmanageable number of candidate rationale sentences. Since this sentence does not contain any of the term patterns we had coded, it does not appear in the results.

The third and most influential reason is that some messages containing actual rationale appear a long time before or after the PEP's decisive state change is committed, as discussed in Section~\ref{sec:rank-ordered}. As our approach includes a heuristic that considers proximity to the commit state, rationale sentences that are temporally more distant from the state change are ranked lower. This is particularly challenging when there are other similar candidate rationale sentences that are closer in time to the state change, as they are likely to have higher $FS$ and thus be ranked higher than the actual rationale sentence.


Evaluating our MBS results using the NDCG metric (as discussed in Section~\ref{sec:effectiveness}) produced an average NDCG value greater than 0.5 for all ranks beyond the top~5 (see Figure~\ref{fig:evaluationmetricsforHBDSS}). For example, for \(k = \textrm{15}\) (i.e., top~15 results) the average NDCG value was 0.6 for \textit{accepted} PEPs and 0.73 for \textit{rejected} PEPs. Intuitively, this means that MBS correctly ranked rationale sentences 60\% of the time for \emph{accepted} PEPs and 73\% of the time for \emph{rejected} PEPs. If we disregard the exact order of results within the top~15, the match is much higher: 82\% and 92\%, respectively (see Table~\ref{tab:reasonsoptimisedmetricsforacceptedandrejected}).

Also, when compared to the prior work of Mrówka \cite{mrowka2015decision} which has identified five rationales in Apache (which they call `approval types'), our work has unearthed six additional ones (a total of 11 rationales). Their work describes the process model for decision-making qualitatively and our work focuses on rationale for decisions quantitatively – e.g., 57 out of 248 were ‘consensus’ decision. In other words, Mrówka’s work identifies ‘consensus’ exists, but does not quantify how often it is used during decision making. In addition, no practical tool was developed by Mrówka as we have pursued in this work.

\subsection{Demonstrating the Practical Utility of our Approach}\label{sec:demonstrating}


We demonstrate the use of our approach to visualize the results from rationale extraction using two examples, both of which are also described in the video mentioned in Section~\ref{sec:GUI_description}. First, we consider the rationale for the acceptance of PEP~572 --- a PEP that was so contentious that the BDFL resigned shortly afterwards. We identified three rationale sentences, of which the most insightful was: ``It's really hard to choose between alternatives, but all things considered I have decided in favor of `NAME := EXPR' instead''\cite{python2018a}, which is an example of \emph{BDFL decree}. SBS ranked this sentence 10th, whereas MBS ranked the corresponding message 8th.

The second example is for the highly contentious PEP~308. The PEP went through several decision-making phases, including a \emph{poll}, a \emph{vote}, and a \emph{complementary vote}, and was subsequently \textit{rejected}. It was eventually \textit{accepted} by the BDFL two years later. Due to the many decision-making phases the PEP went through, the rationale behind its eventual acceptance were stated in a single message: the PEP~308 summary message~(\cite{python2005d}). SBS ranked two of these rationale-containing sentences 8th and 9th, whereas MBS ranked the corresponding message 2nd. 

We also found some surprising results in our evaluation. There were four instances where the rationale sentence captured by Rationale Miner was better than what we had captured manually. For example, the following sentence was manually flagged as a rationale-containing sentence for PEP 465, and was ranked 26th by SBS: ``Because this way is compatible with the existing consensus, and because it gives us a consistent rule that all the built in numeric operators also apply in an elementwise manner to arrays; the reverse convention would lead to more special cases.'' However, a more insightful sentence, which provided more details of the ``consensus'' was ranked 7th by SBS: ``the result is a strong consensus among both numpy developers and developers of downstream packages that numpy.matrix should essentially never be used because of the problems caused by having conflicting duck types for arrays.'' Both sentences were from the same message \cite{python2014a}, which MBS ranked 5th. While these surprises were observed in four instances, one could argue these four cases were due to human error (i.e., the human coder failed to label a rationale sentence). This highlights that our system indeed has found those sentences correctly. We believe this also shows the effectiveness of our approach to unearthing rationale.

\subsection{Usage of Rationale extraction tool}\label{sec:intendeduser}

The intended users for the heuristic-based rationale extraction tool are both practitioners and researchers. \emph{Practitioners} (developers and decision-makers in Python) can use the tool to unearth rationale (e.g., 11 types of rationale such as lazy consensus, with some of them unknown to stakeholders such as ‘rough consensus’), and quantify their uptake (e.g., how often is rough consensus used?) Also, they can use the system to understand when, why, and how decisions were made through retrospective analysis (since the details of these discussions are not available in the PEP documents themselves). The tool can be used for comparative studies across communities. 

\emph{Researchers} can use the tool to examine the nature of the rationale for decisions (similar to what we have presented in Figure \ref{fig:dmconceptspiechart}). Also, several research questions can be answered subsequently. To name a few: a) what are the nature of the arguments for and against acceptance (e.g. strong vs. weak, or evidence-driven or anecdotal rationale) b) what are the ‘signatures’ of success (what facilitates the PEP acceptance?), and c) how can bad PEPs (e.g. inept PEPs) be avoided at the outset. Thus, our work will spur further investigations.

Extending beyond the Python community, our findings also has implications regarding decision making processes in other open source projects. These include: a) reuse of our methodological approach (i.e, 13 heuristics) as a template towards rationale extraction in other OSS communities, b) knowledge of the existence of more rationale (11 in our work compared to five in Mrówka) c) learning from why PEPs fail (e.g., inept PEPs), and then use that to avoid failures in other communities (e.g., OpenJDK).

\subsection{Threats to Validity}
In this section, we consider threats to construct validity, internal validity, and external validity. \emph{Threats to construct validity} refer to the appropriateness of our evaluation measures. In our scoring construct we used 13 different heuristics based on prior work and our experience with manual analysis of emails. It is possible that we may have missed other heuristics that may be influential in identifying rationale sentences. 
For formal evaluation of rationale sentences we considered the NDCG metric which has been previously used to evaluate ranking of emails. Thus, the threats to the use of this construct are minimal. 

A \emph{threat to the internal validity} of the study is that the labelling process (i.e., labelling a sentence as a rationale-containing sentence or not), is a subjective one and there might be errors. To overcome this threat, the second author of the paper double-checked the labels of the 300 rationale sentences, and any discrepancy was resolved by discussion. Another threat to the internal validity is that we may have missed some rationale containing sentences where the human language is ambiguous. However, we believe this threat is somewhat mitigated because almost all the rationale we have identified have been identified in different literature on decision-making, except for rough consensus. 

A \emph{threat to external validity} is the generalization of our approach to extract rationale. We concede that our approach is applicable only to software development communities that have decision-making processes similar to Python. For example the rationale for decisions in Java Community that follows Java Enhancement Proposals (JEPs) that were inspired by PEPs. However, we believe the approach employed in the Rationale Miner can be extended to extract rationale from other communities such as the Linux and Perl projects.

\section{Conclusion}\label{sec:conclusion}

There has been relatively little prior research in software engineering investigating automated extraction of rationale behind how decisions are made. 
The current work bridges this gap by proposing a heuristics-based approach that is operationalized in the Rationale Miner tool. Two approaches to rationale extraction (sentence-based and message-based) were proposed and evaluated, using the Python OSSD project as a case study. 

Our results identified 11 different ways of reaching a decision outcome. 
Rationale Miner was able to correctly rank rationale-containing sentences for PEP decisions into the top~10 results nearly two-thirds of the time for both \emph{accepted} and \emph{rejected} PEPs. Our approach can be extended and used in other OSSD communities that have decision-making structures similar to that of Python, such as the Java Community Process with its JEPs. 
The heuristics may need to be remodelled to accommodate any additional factors in these communities.

\section*{Acknowledgment}


The first author thanks Russell Education Trust for their financial support to pursue a part of his PhD study.

\bibliographystyle{IEEEtran}
\bibliography{reasons}


\end{document}